\documentclass[twocolumn,amssymb,amsmath,aps,pre,reprint,floatfix]{revtex4-2}
\usepackage{hyperref}
\usepackage{graphicx}
\graphicspath{{figs/}}
\begin{document}

\title{Invariant percolation properties in random isotropic systems of conductive discorectangles on a plane: From disks to sticks}%

\author{Yuri~Yu.~Tarasevich}
\email[Corresponding author: ]{tarasevich@asu.edu.ru}

\author{Andrei~V.~Eserkepov}
\email{dantealigjery49@gmail.com}

\affiliation{Laboratory of Mathematical Modeling, Astrakhan State University, Astrakhan 414056, Russia}

\date{\today}%
\begin{abstract}
Recently, some  eccentricity-invariant properties of random, isotropic, two-dimensional (2D) systems of conductive ellipses have been reported [\href{https://doi.org/10.1103/PhysRevB.104.184205}{Phys. Rev. B \bf{104}, 184205 (2021)}].  Moreover, the authors suggested that this invariance might also be observed in systems with other particle geometries having zero-width sticks as the limiting case. To check this suggestion, we studied 2D random systems of isotropically-placed, overlapping, identical discorectangles (stadia) with aspect ratios ranging from 1 (disks) to $\infty$ (zero-width sticks). We analyzed the effect of the aspect ratio and the number density of conductive discorectangles on the behavior of the electrical conductivity, the local conductivity exponent, and the current-carrying backbone. Our own computer simulations demonstrate that some of the properties of random, isotropic 2D systems of conductive discorectangles are insensitive to the aspect ratios of the particles.
\end{abstract}
\maketitle

\section{Introduction}\label{sec:intro}
The 2D systems of randomly-placed, metallic nanowires and nanorods are being extensively studied. The interest in these systems is inspired by their combination of high electrical conductivity with excellent optical transparency that is in demand in numerous technological applications~\cite{Gao2016,Sohn2019} such as touch screens~\cite{Nguyen2022}, transparent heaters~\cite{Gupta2016}, solar cells~\cite{Zhang2020}, and flexible electronics~\cite{Li2020}.

To mimic the shape of elongated particles and, at the same time, simplify the simulations, different simple geometrical figures are used, e.g., zero-width sticks (rods)~\cite{Mertens2012,Chatterjee2013,Tarasevich2018}, rectangles~\cite{Li2013}, ellipses~\cite{Li2016,Alvarez2021}, superellipses~\cite{Lin2019}, and discorectangles (stadia)~\cite{Tarasevich2020}. Comparisons of some of the properties of random 2D systems of the above particles have been collected in Ref.~\cite{Lebovka2020}. The electrical properties of the 2D systems of randomly-placed conductive particles of the above shapes have also been studied~\cite{Alvarez2021,Zezelj2012,Aryanfar2021,Tarasevich2022PCCP}, with the greatest attention being paid to the simplest shape, i.e., to zero-width sticks~\cite{Balberg1983,OCallaghan2016,Forro2018}, and the particular case when only the junction resistance is taken into account~\cite{Kim2019}.

Some  eccentricity-invariant properties of random isotropic, two-dimensional (2D) systems of conductive ellipses have recently been reported~\cite{Alvarez2021}. The authors have suggested that this invariance might also be observed in systems with other particle geometries having the zero-width sticks as the limiting case. To check this conjecture, we studied 2D random isotropic systems of conductive overlapping discorectangles. Their aspect ratios ranged from 1 (disks) to $\infty$ (zero-width sticks). We analyzed the behavior of the electrical conductivity, the local conductivity exponent, and the backbone in respect of the aspect ratio and the number density of these conductive discorectangles. We have compared our results with the published results for ellipses~\cite{Alvarez2021} and sticks~\cite{Zezelj2012}.

The rest of the paper is constructed as follows. Section~\ref{sec:methods} describes some technical details of our simulation. In Section~\ref{sec:results}, we present our main results and discuss some open questions. Section~\ref{sec:concl} summarizes the main results and suggests possible directions for further study.

\section{Methods}\label{sec:methods}
\subsection{Sampling}
\begin{figure}[hbt]
  \centering
  \includegraphics[width=0.4\columnwidth]{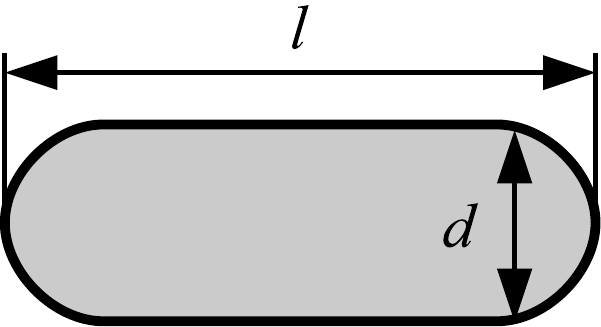}\\
  \caption{Discorectangle (stadium).\label{fig:stadium}}
\end{figure}

A discorectangle (a stadium) is a rectangle with semicircles at a pair of opposite sides (Fig.~\ref{fig:stadium}). Its aspect ratio is
\begin{equation}\label{eq:AR}
  \varepsilon = \frac{l}{d}.
\end{equation}
When $\varepsilon = 1$, a discorectangle reduces in a disk. The limiting case $\varepsilon = \infty$ corresponds to a zero-width (widthless) stick.

\begin{figure}[!hbt]
  \centering
  \includegraphics[width=0.5\columnwidth]{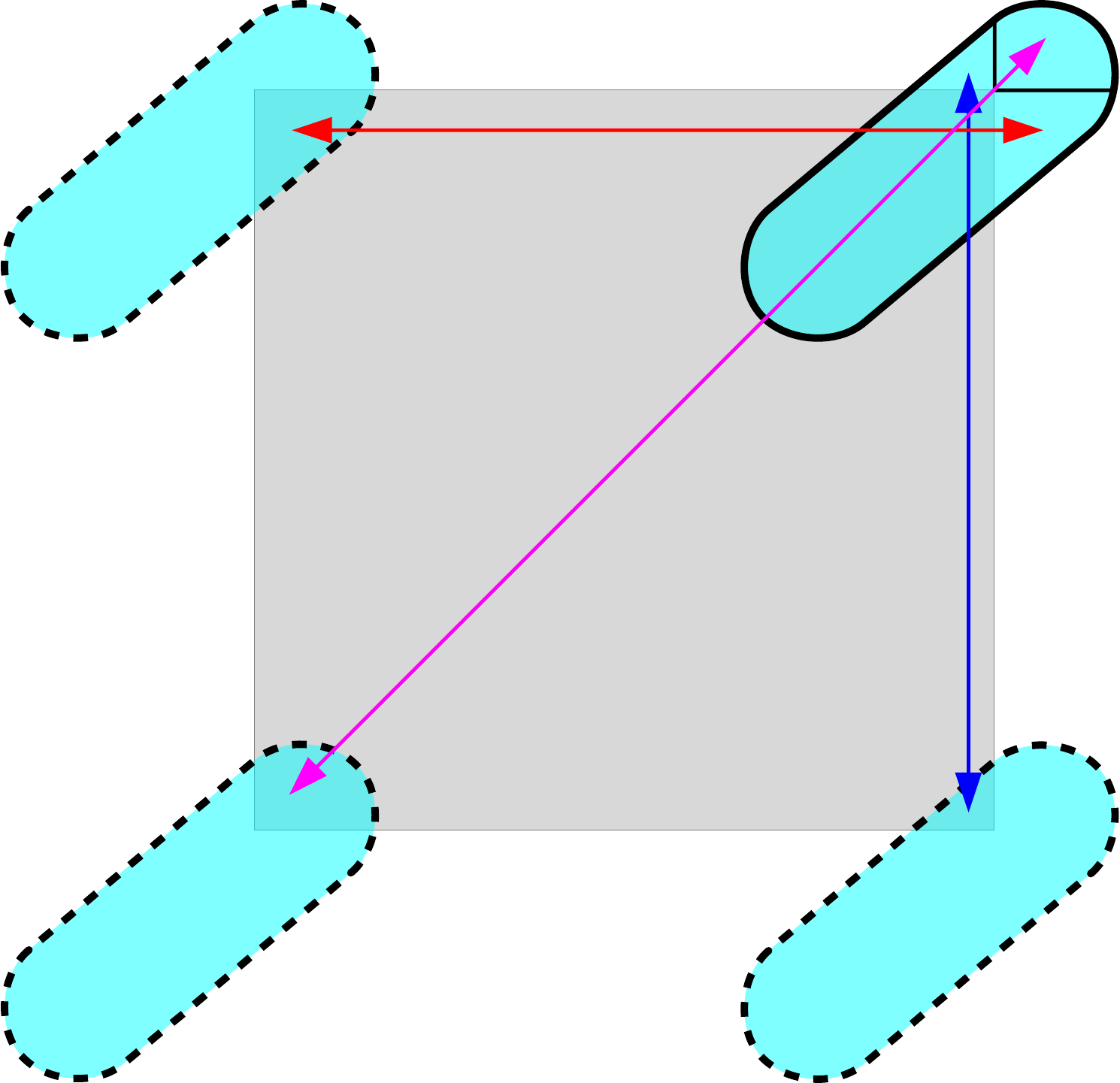}\\
  \caption{Example of the effect of PBCs for the most complicated case. The discorectangle in the top right corner extends beyond the square domain $L \times L$. Due to the PBCs, the parts of the particle that are outside the domain are duplicated in the remaining corners of the domain (these ``ghost'' particles are indicated using dashed lines).\label{fig:PBC}}
\end{figure}

Basically, in our study we used discorectangles with four alternative values of the aspect ratio, viz., $\varepsilon = 1,7,20,\infty$, while the discorectangle's length was fixed, $l=1$. Some additional investigations have been performed for intermediate values of $\varepsilon$. Identical, permeable discorectangles with the chosen value of aspect ratio were randomly placed on a substrate. Their centers were independent and identically distributed within a square domain of size $L \times L$, while their orientations were equiprobable. To reduce the finite-size effect, periodic boundary conditions (PBCs) were applied along both mutually perpendicular directions (Fig.~\ref{fig:PBC}).

The number density of the deposited particles is the number of particles $N$ per unit area, i.e.,
\begin{equation}\label{eq:ND}
  n = \frac{N}{L^2}.
\end{equation}
Another widely used quantity to characterize a deposit is the total area fraction of the deposited particles,
\begin{equation}\label{eq:eta}
  \eta = A n,
\end{equation}
where $A$ is the area of each particle. In contrast to open boundary conditions, PBCs ensure that the relation~\eqref{eq:eta} is exact. In the case of elongated particles, PBCs ensure the isotropic deposition of such particles, while the closed boundary conditions force the particles to align along the boundaries, which leads to locally anisotropic systems.

\subsection{Overlapping of the deposited particles}
\begin{figure}[!htb]
  \centering
  \includegraphics[width=\columnwidth]{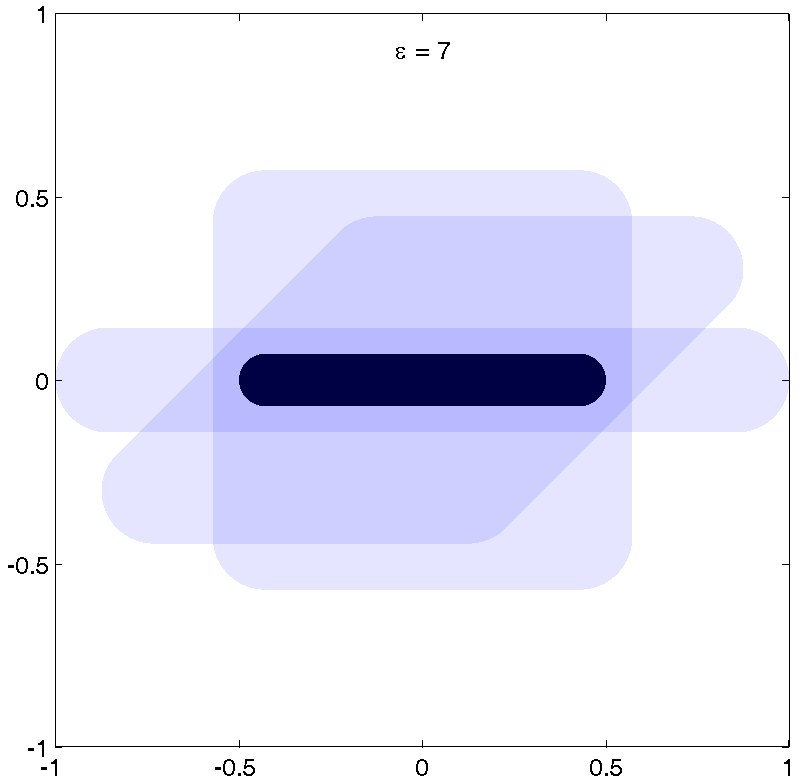}
  \caption{Examples of the excluded areas of a discorectangle for different angles between this discorectangle and a second one; $\vartheta = 0, \pi/4, \pi/2$,  $\varepsilon = 7$.\label{fig:Aex7}}
\end{figure}

Two discorectangles are considered as connected if they overlap. The overlapping occurs when the center of the second discorectangle is located within the excluded area of the first one~\cite{Balberg1984}. The excluded area depends on the mutual orientation of the discorectangles
\begin{equation}\label{eq:Aex-angle}
  A_\text{ex} = \sin\vartheta (l-d)^2  + 4 d (l-d) +  \pi d^2,
\end{equation}
where $\vartheta$ is the angle between the two discorectangles~\cite{Balberg1984,Volkov2020} (Fig.~\ref{fig:Aex7}).

\begin{figure}[!htb]
  \centering
  \includegraphics[width=\columnwidth]{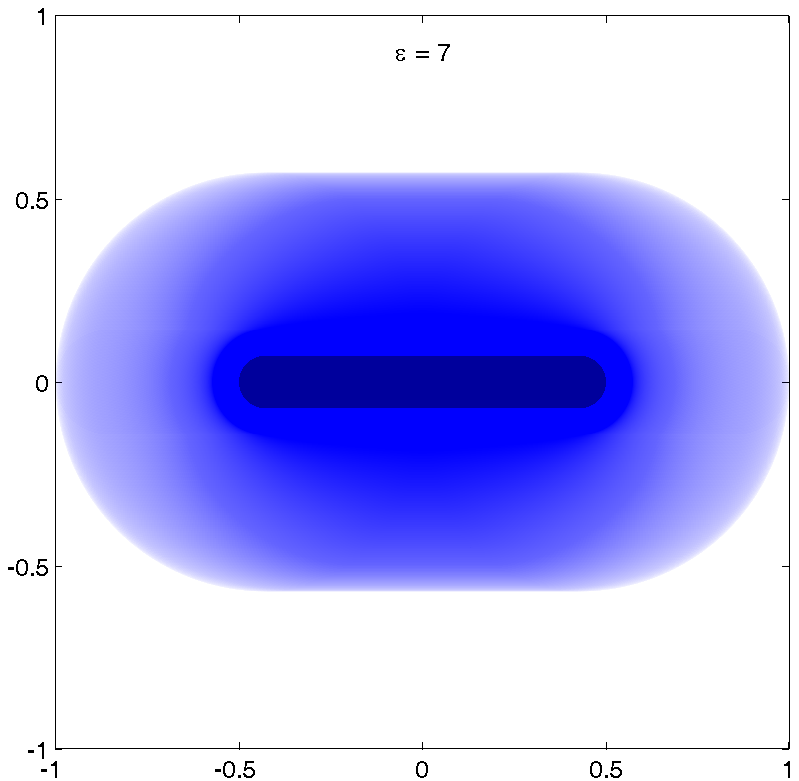}
  \caption{Example of the probability map when all orientations of the discorectangles are equiprobable (isotropic deposition of discorectangles) for $\varepsilon = 7$. The reference discorectangle is shown. The darkest shade corresponds to the probability 1, i.e., when the center of the second discorectangle is located within this area, and overlapping of these two discorectangles is ensured.\label{fig:AR7}}
\end{figure}

Superposition of the excluded areas for all possible mutual angles between two discorectangles can be treated as a probability map, i.e., the probability that a randomly oriented discorectangle will overlap an original discorectangle. Figure~\ref{fig:AR7} presents the probability map when all orientations are equiprobable (isotropic deposition).

\begin{figure}[hbt]
  \centering
  \includegraphics[width=0.9\columnwidth]{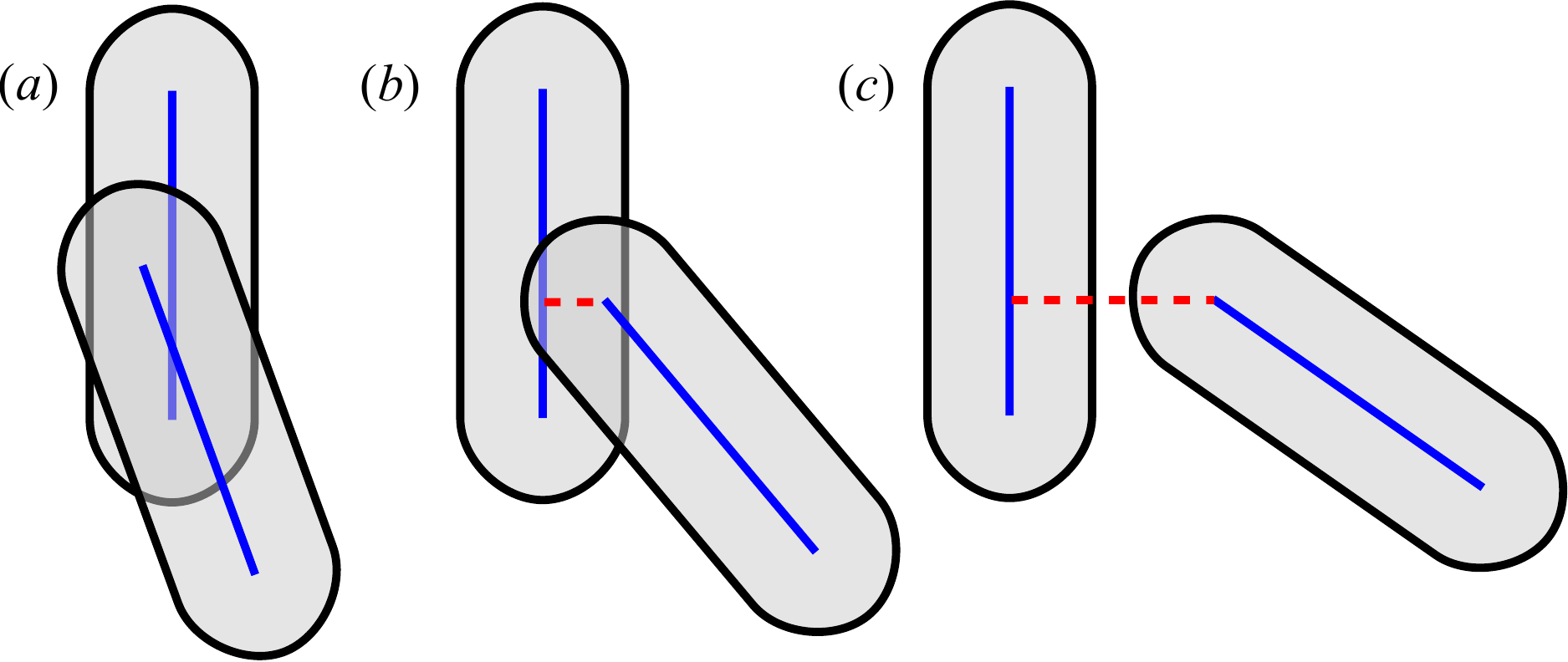}\\
  \caption{(a) The discorectangles overlap, since their midlines intersect each other.
  (b) The discorectangles overlap, since the shortest distance between their midlines is less than the width of the discorectangle, $d$. (c) The corresponding discorectangles do not overlap, since  the shortest distance between their midlines is larger than the width,~$d$, of a discorectangle.\label{fig:intersection}}
\end{figure}

To detect any overlapping of the discorectangles, we used their midlines (Fig.~\ref{fig:intersection}).
\begin{enumerate}
  \item The discorectangles overlap, if their midlines intersect each other [Fig.~\ref{fig:intersection}(a)].
  \item If the two midlines do not intersect each other, but the shortest distance between these midlines is less than the width of the discorectangle, $d$, the corresponding discorectangles overlap [Fig.~\ref{fig:intersection}(b)].
  \item If the two midlines do not intersect each other and the shortest distance between these midlines is larger than the width of the discorectangle, $d$, the corresponding discorectangles do not overlap [Fig.~\ref{fig:intersection}(c)].
\end{enumerate}
The limiting cases are obvious. To check whether two zero-width sticks intersect, consideration of just the first item above is enough. The two disks intersect each other when the distance between their centers is less than~$d$.

The probability that two identical particles intersect with each other can be found using the excluded area concept~\cite{Balberg1984}
\begin{equation}\label{eq:P}
  P = \frac{A_\text{ex}}{L^2},
\end{equation}
where, in the case of isotropically-placed discorectangles, the angle-averaged excluded area is
\begin{equation}\label{eq:Aex}
  \langle A_\text{ex}\rangle = \frac{2}{\pi} (l-d)^2  + 4 d (l-d) +  \pi d^2
\end{equation}
(see, e.g., Refs.~\onlinecite{Balberg1984,Volkov2020})
or
\begin{equation}\label{eq:AexAR}
\langle A_\text{ex}\rangle = l^2\left[\frac{2}{\pi} {\left(1-\varepsilon^{-1}\right)}^2 + 4\varepsilon^{-1} \left(1-\varepsilon^{-1}\right) + \pi \varepsilon^{-2} \right]
\end{equation}
(see, e.g., Ref.~\onlinecite{Lebovka2020}).
In the case of disks ($l=d$, $\varepsilon =1$),
\begin{equation}\label{eq:Aexdisk}
 \langle A_\text{ex} \rangle = \pi d^2 = \pi l^2,
\end{equation}
while in the case of zero-width sticks ($d=0$, $\varepsilon = \infty$),
\begin{equation}\label{eq:Aexstick}
 \langle A_\text{ex} \rangle = \frac{2 l^2}{\pi}.
\end{equation}
The mean number of intersections per deposited particle is $\langle k \rangle = (N - 1) P$ or
\begin{equation}\label{eq:kmean}
\langle k \rangle \approx n P L^2,
\end{equation}
when $N \gg 1$. Hence, the total number of contacts (junctions) between particles is
\begin{equation}\label{eq:Nj}
  N_\text{j} = N \frac{n P L^2}{2} =  \frac{n^2 L^2 \langle A_\text{ex}\rangle}{2},
\end{equation}
while the number density of contacts is
\begin{equation}\label{eq:nj}
  n_\text{j} =  \frac{n^2 \langle A_\text{ex}\rangle}{2}.
\end{equation}
Formula~\eqref{eq:nj} is valid for identical particles of any shape, not only for discorectangles. Moreover, formula~\eqref{eq:nj} is valid for anisotropic systems when the angle-averaged excluded area is calculated using the appropriate angle-distribution function. Thereby, there is a quadratic dependency of the number density of the contacts between the deposited particles on the number density of the deposited particles.

\begin{figure}[hbt]
  \centering
  \includegraphics[width=0.9\columnwidth]{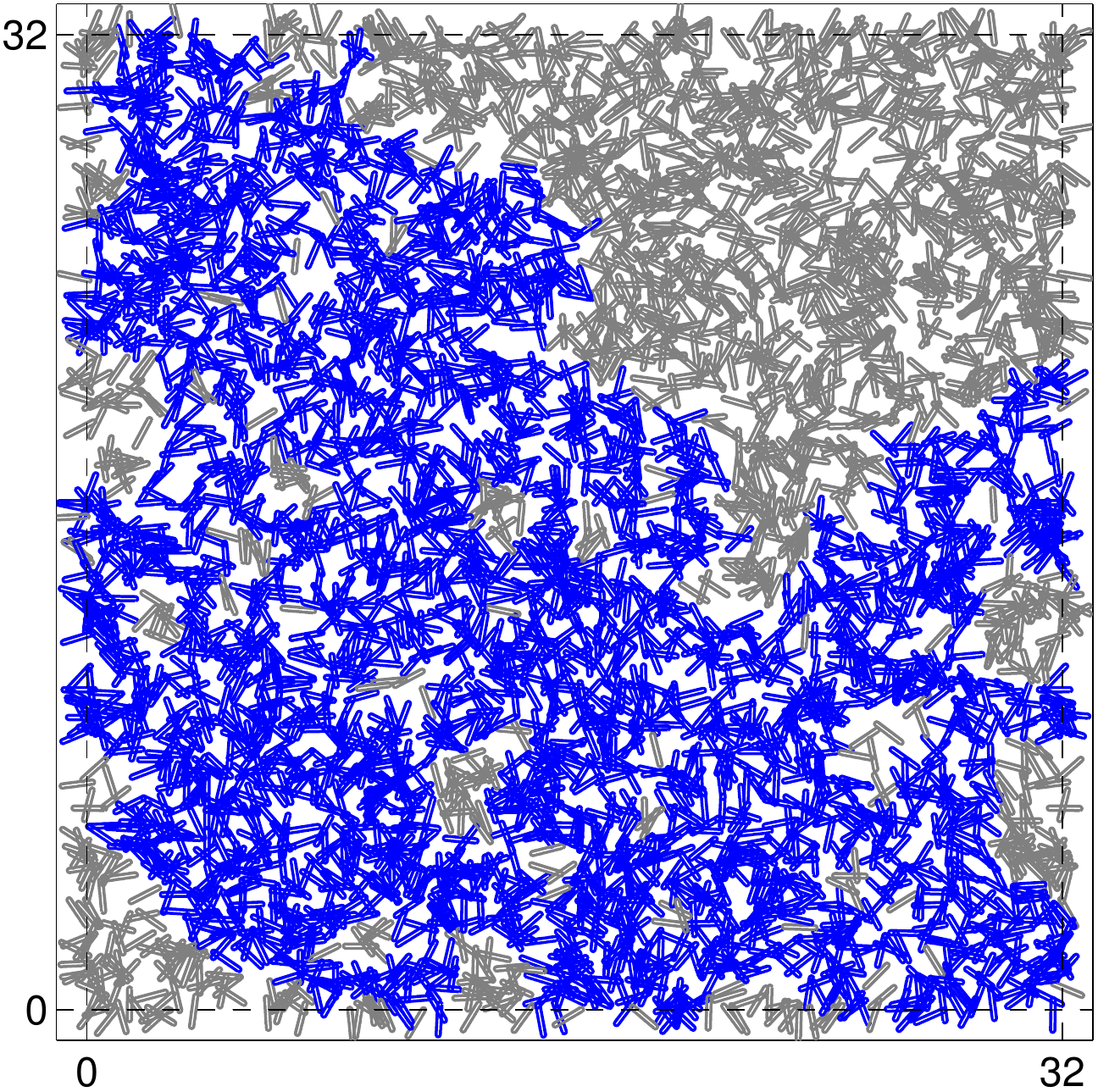}\\
  \caption{Example of a system under consideration exactly at the percolation threshold. The spanning cluster is highlighted. The aspect ratio of the discorectangles is 7.\label{fig:system}}
\end{figure}

For all values of discorectangle aspect ratio, we used domains of a fixed size $L=32 l$. To efficiently determine the percolation threshold (occurrence of a percolation cluster that spans the system in a given direction), the union-find algorithm~\cite{Newman2000PRL,Newman2001PRE} was used. Figure~\ref{fig:system} exhibits an example of a system under consideration exactly at the percolation threshold. The spanning cluster is highlighted. The aspect ratio of the discorectangles is~7.

We used the normalized number density of the deposited particles, viz., $n/n_\text{c} - 1$, in all our figures. Here, $n_\text{c}$ is the percolation threshold. Notice that the normalized number density is equal to the normalized total area fraction $\eta/\eta_\text{c} - 1$.

\subsection{Electrical properties}
To study the electrical properties of the deposit, we added superconductive busses to the two opposite boundaries of the system under consideration. Only the resistance of junctions between particles was taken into account. This assumption of junction resistance dominance has previously been widely used~\cite{Zezelj2012,Kim2019,Alvarez2021}. In such a way, the system under consideration can be transformed into a network, in which the edges correspond to the contacts (overlaps) between deposited particles, while the vertices of the network correspond to the deposited particles. This network is not planar. In it, each edge represents the resistance $R_\text{j}$ (the conductance is $\sigma_\text{j}$), i.e., a random resistor network (RRN) is being considered. To be precise, there is an irregular network with identical branch resistances. For this particular case of the RRN, some analytical results derived on the basis of the Foster theorem~\cite{Foster1949,Foster1961} are known~\cite{Marchant1979}.

Let $y_i$ be the admittance associated with the $i$-th branch, while $Y_i$ is the admittance seen from the endpoints of the $i$-th branch when $y_i$ is disconnected. For an irregular network with identical branch admittances $y_i = y_m$,
  \begin{equation}\label{eq:Marchant7}
  \left\langle\frac{y_m}{y_m + Y_i}\right\rangle = \frac{2}{\langle{\deg V}\rangle },
  \end{equation}
where $\langle \deg V \rangle$ is the average degree of the network nodes~\cite{Marchant1979}. In our case, accounting for Eqs.~\eqref{eq:kmean} and \eqref{eq:P}, $\deg V = \langle k \rangle  \approx n A_\text{ex}$
  \begin{equation}\label{eq:MarchantDR}
  \left\langle\frac{\sigma_\text{j}}{\sigma_\text{j} + Y_i}\right\rangle = \frac{2}{n  A_\text{ex}}.
  \end{equation}

To identify a current-carrying part (the backbone) of the percolation cluster, we used the algorithm as follows (for the sake of clarity some irrelevant details have been omitted).

Let $G'$ be the percolation cluster of the network $G$. The vertices of $G'$ belonging to one bus are considered as inputs, while the vertices belonging to the other bus are considered as outputs. Initially, all vertices and all edges of the network $G'$ are marked as ``unremoved''.
\begin{enumerate}
  \item Add vertices $V_1$ and $V_2$ to the network $G'$ in such a way that $V_1$ is connected to all the inputs, while $V_2$ is connected to all the outputs.
  \item Find all articulation points.
  \item Check each articulation point.
  If the current  articulation point (vertex $X$) is marked as ``unremoved'', then
  \begin{enumerate}
    \item Find all vertices adjacent to $X$.
    \item Mark $X$ and all its incident edges as ``removed''.
    \item Check each vertex, $Y$, that is adjacent to $X$.
    \begin{enumerate}
      \item Check the presence of paths from vertex $Y$ to vertices $V_1$ and $V_2$ in the subgraph $H$, which consists of ``unremoved'' vertices and edges of the network $G'$.
      \item If there is a path leading from vertex $Y$ to neither $V_1$ nor $V_2$, then we mark as ``removed'' all vertices and edges of the connected component of network $H$ containing vertex $Y$.
      \end{enumerate}
    \end{enumerate}
    \item Mark the vertex $X$ as ``unremoved''.
        \item Mark as ``unremoved'' all edges between the vertex $X$ and the vertices marked as ''unremoved''.
  \end{enumerate}
In fact, we are looking for a geometrical backbone, i.e., a biconnected component of the network. A geometrical backbone can contain perfectly balanced bonds (Wheatstone bridges). Since the potential
difference between the ends of a perfectly balanced bond is equal to zero, electrical current through this bond is absent~\cite{Li2007}. However, it is intuitively clear that the fraction of  perfectly balanced bonds has to be negligible, if there are any at all. By contrast, direct identification of the current-carrying part of the percolation cluster is hardly reliable, since some apparent, but actually non-existent, currents may arise both in dead ends and in perfectly balanced bonds due to rounding-off errors. These currents may be of the same order of magnitude as the real currents in some parts of the network.

A definition for the local transport exponent~\cite{Bernasconi1978,Grimaldi2006,Zezelj2012,Alvarez2021} is
\begin{equation}\label{eq:tlocal}
t = \frac{\mathrm{d} \ln \sigma}{\mathrm{d} \ln (\eta - \eta_\text{c})} = \frac{\eta - \eta_\text{c}}{\sigma} \frac{\mathrm{d} \sigma}{\mathrm{d}\eta}= \frac{n - n_\text{c}}{\sigma} \frac{\mathrm{d} \sigma}{\mathrm{d}n}.
\end{equation}
Using an analytical formula for the electrical conductivity of 2D system of randomly placed conductive sticks that was obtained within a mean-field approach~\cite{Tarasevich2022PCCP}
\begin{equation}\label{eq:MFAsigmaJDRcont}
\sigma = \frac{n^2 l^4 }{12 \pi R_\text{j}},
\end{equation}
the local transport exponent may be derived as
\begin{equation}\label{eq:tMFAsticks}
t = 2\frac{n - n_\text{c}}{n}.
\end{equation}
This local transport exponent tends to 2, when $n \gg  n_\text{c}$.

The error bars in the figures correspond to the standard deviation of the mean. When
not shown explicitly, they are of the order of the marker size.

\section{Results}\label{sec:results}
\begin{figure}[!htb]
  \centering
  \includegraphics[width=\columnwidth]{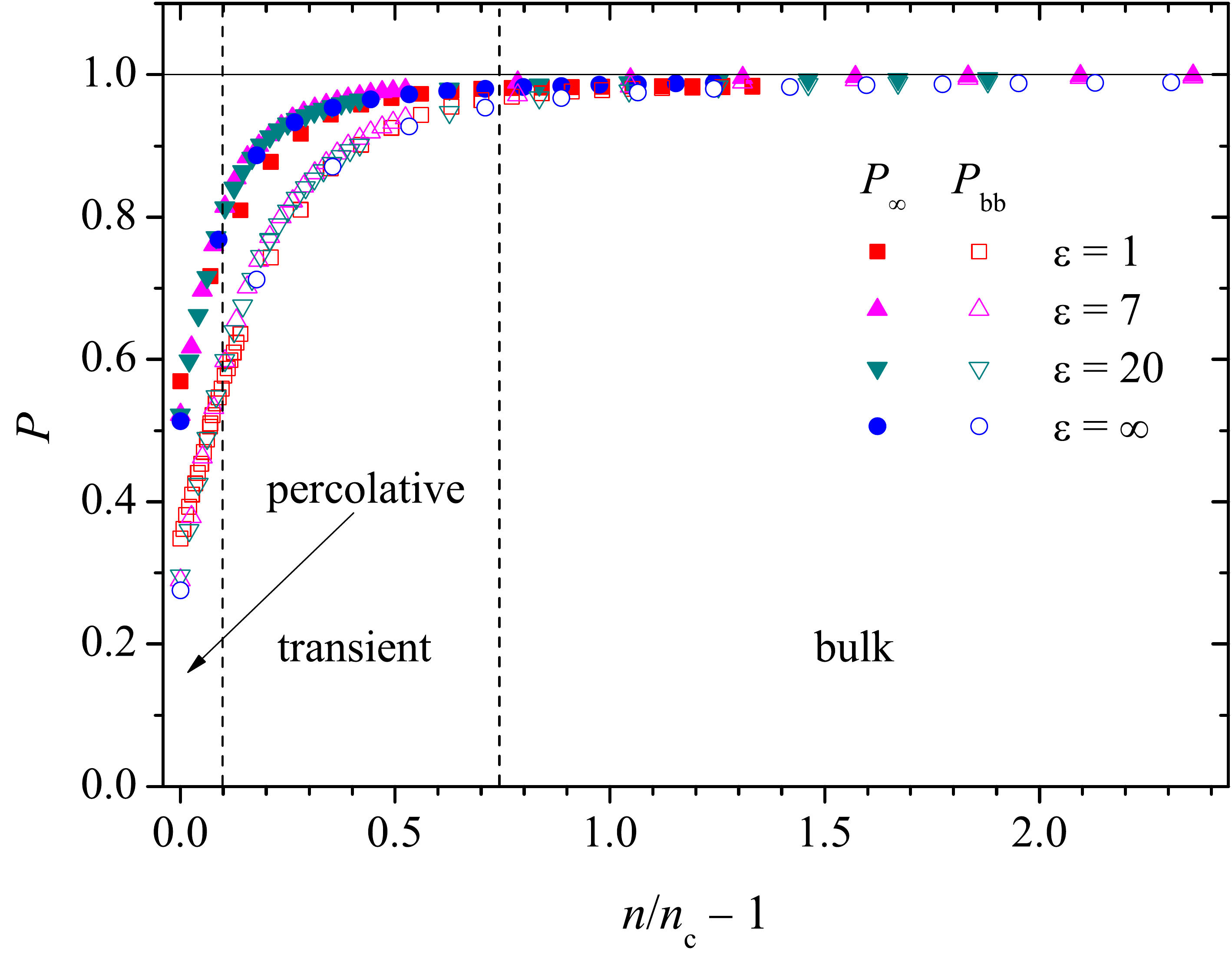}
  \caption{Strength of the percolation cluster (filled markers) and its backbone (open markers) against the normalized number density, $n/n_\text{c} - 1$, for discorectangles having different aspect ratios. The results are averaged over 1000 independent runs.}\label{fig:strength}
\end{figure}

Figure~\ref{fig:strength} demonstrates the strength of the percolation cluster (filled markers) and its backbone (open markers) against the normalized number density, $n/n_\text{c} - 1$, for  discorectangles possessing different aspect ratios, i.e., from disks to sticks. First of all, the results seem to be independent or almost independent of the aspect ratio. However, a small deviation for disks can be noticed. This deviation requires an additional detailed study. Moreover, three different regimes can be observed, viz., (i) a percolation regime ($n \lessapprox 1.1 n_\text{c} $), (ii) a transient regime ($n_\text{c} \lessapprox n \lessapprox 1.7n_\text{c}$), and (iii) a bulk regime ($n \gtrapprox 1.7n_\text{c}$). In the bulk regime, all, or almost all, the particles belong to the percolation cluster. It is noteworthy that universal (aspect ratio invariant) behavior is observed in all three regimes.

\begin{figure}[!htb]
  \centering
  \includegraphics[width=\columnwidth]{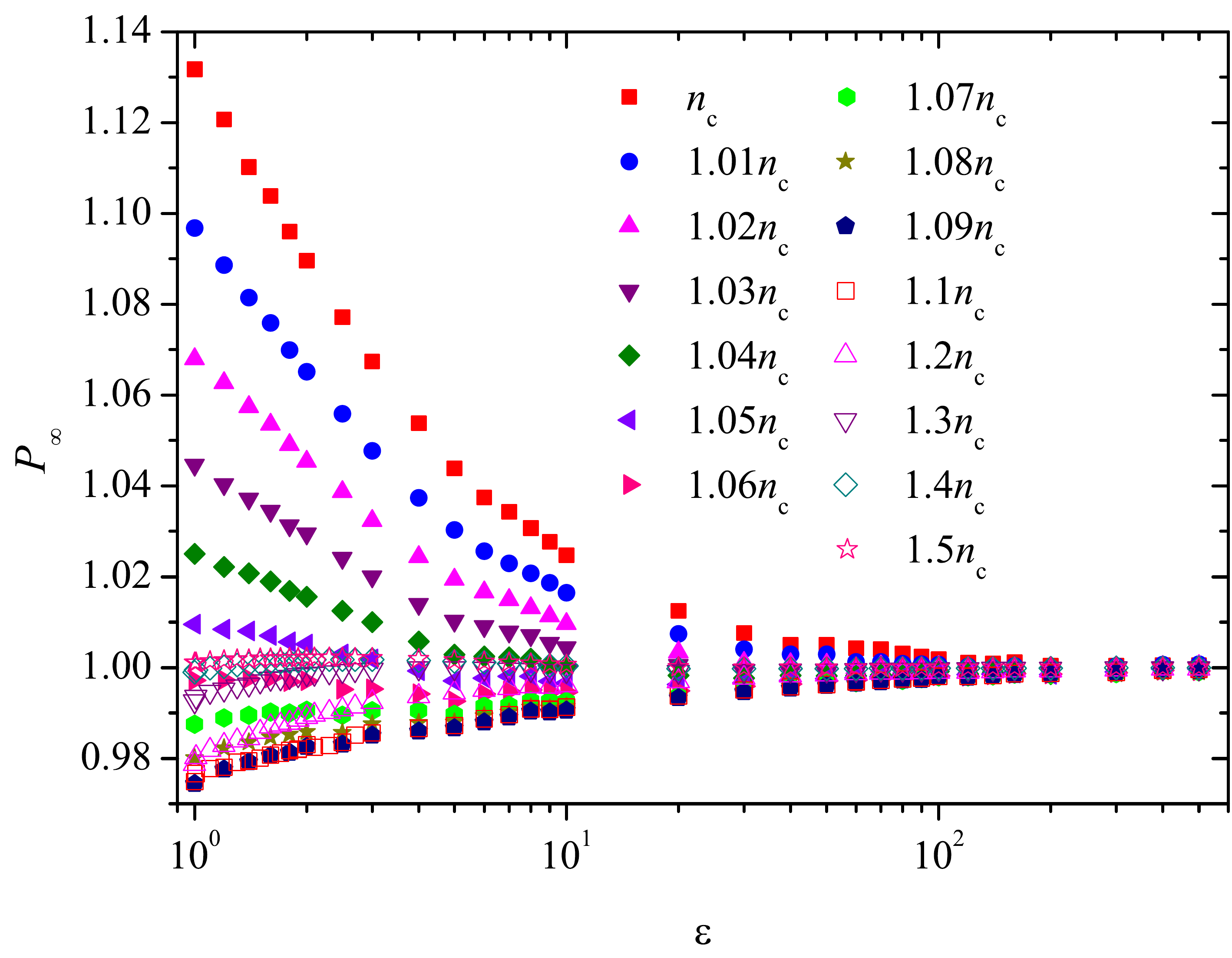}
  \caption{Strength of the percolation cluster versus the aspect ratio for different values of the number density. The results are averaged over 100\,000 independent runs.}\label{fig:compar}
\end{figure}

Figure~\ref{fig:compar} presents a close look at the behavior of the percolation cluster strength. The strength of the percolation cluster is normalized by the strength of the percolation cluster of the sticks ($\varepsilon = \infty$). For any studied values of the number density, this normalized strength of the  percolation cluster approaches unity as the aspect ratio increases.

\begin{figure}[!htb]
  \centering
  \includegraphics[width=\columnwidth]{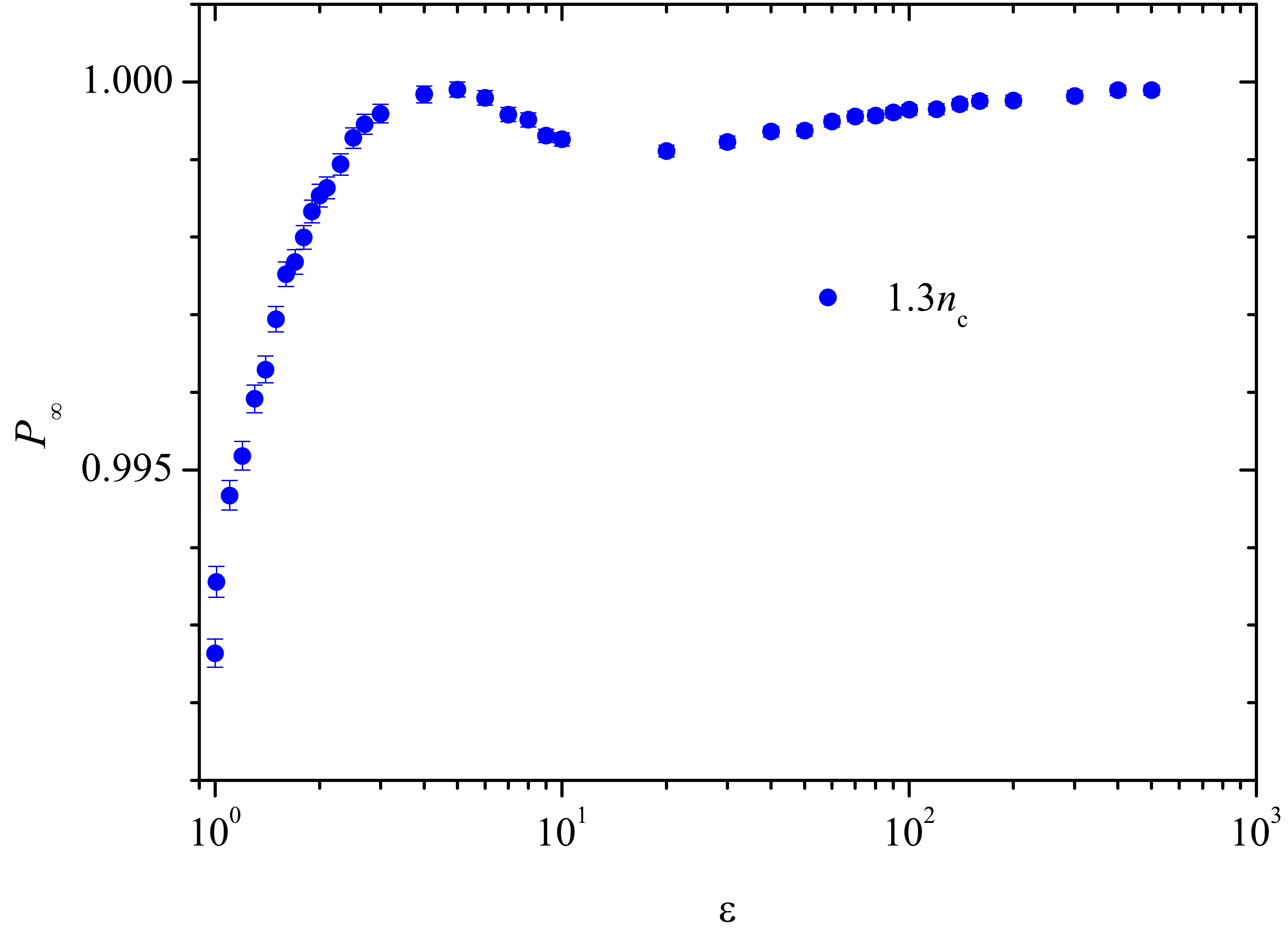}
  \caption{Strength of the percolation cluster on the aspect ratio for $n=1.3n_\text{c}$. The results are averaged over 100\,000 independent runs.}\label{fig:Pinf03}
\end{figure}

However, this approach may be nonmonotonic (Fig.~\ref{fig:Pinf03}).

\begin{figure}[!htb]
  \centering
  \includegraphics[width=\columnwidth]{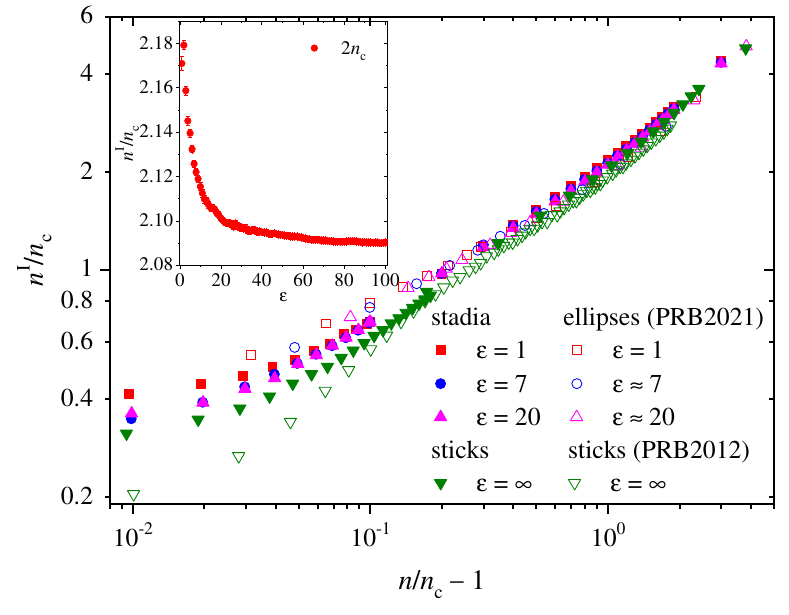}
  \caption{Density of particles belonging to the backbone of the percolation cluster, normalized by the density of particles at the percolation threshold, $n_c$. Filled markers correspond to our results, while open markers correspond to data extracted from the literature. PRB2012 refers to  Ref.~\onlinecite{Zezelj2012}, PRB2021 refers to Ref.~\onlinecite{Alvarez2021}. Inset: dependence of the normalized density of the particles belonging to the backbone of the percolation cluster on the aspect ratio of the discorectangle for the fixed value of the number density of the particles of $n = 2n_c$. Our results are averaged over 100 independent runs.}\label{fig:particles}
\end{figure}

Figure~\ref{fig:particles} shows the density of particles belonging to the backbone of the percolation cluster, normalized by the density of particles at the percolation threshold, $n_c$. The filled markers correspond to our results, while the open markers correspond to data extracted from the literature, viz., the data for ellipses (PRB2021) have been adapted from Ref.~\onlinecite{Alvarez2021}, while the data for sticks (PRB2012) have been adapted from Ref.~\onlinecite{Zezelj2012}. A noticeable divergence when $n/n_c - 1 \lessapprox 0.2$ may arise due to both a finite-size effect and a strong dependency on the accuracy of the percolation threshold estimate. In any case, for $n/n_c - 1 \gtrapprox 0.2$, all data collapse to one curve, confirming the proposition about an aspect-ratio-invariant behavior~\cite{Alvarez2021}. The inset shows a slight monotonic decrease in the normalized density of the particles belonging to the backbone of the percolation cluster with an increase in the value of the aspect ratio of the discorectangles for the fixed value of the number density of the particles of $n = 2n_c$. Our results are averaged over 100 independent runs.

\begin{figure}[!htb]
  \centering
  \includegraphics[width=\columnwidth]{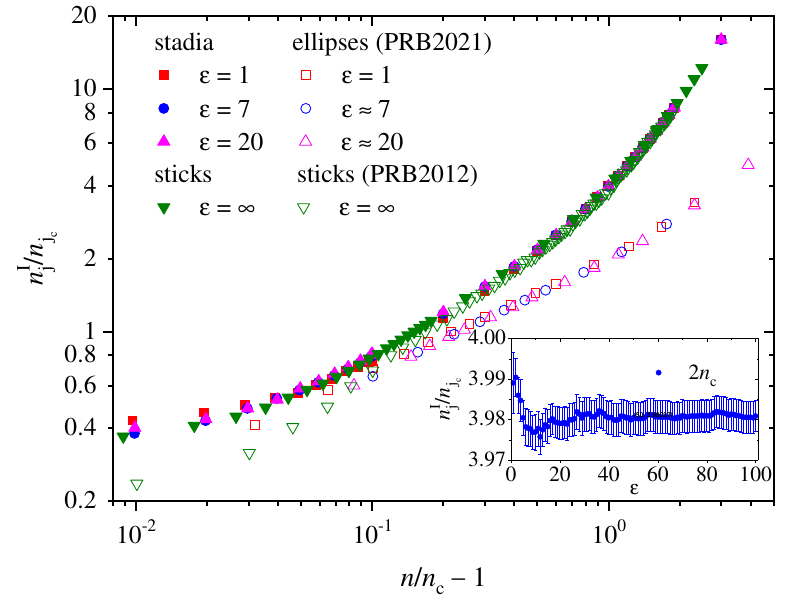}
  \caption{Density of junctions or bonds in the backbone normalized by the density of junctions at the percolation threshold, $n_{j_c}$. Filled markers correspond to our results, while open markers correspond to data extracted from the literature. PRB2012 refers to  Ref.~\onlinecite{Zezelj2012}, PRB2021 refers to Ref.~\onlinecite{Alvarez2021}. Inset: dependence of the normalized density of junctions belonging to the backbone of the percolation cluster on the aspect ratio of the discorectangles for the fixed value of the number density of the particles, $n = 2n_c$. Our results are averaged over 100 independent runs.}\label{fig:junctions}
\end{figure}

Figure~\ref{fig:junctions} demonstrates the density of the junctions or bonds in the backbone normalized by the density of the junctions at the percolation threshold, $n_{j_c}$. The filled  markers correspond to our results, while the open markers correspond to data extracted from the literature, viz., the data for ellipses (PRB2021) have been adapted from Ref.~\onlinecite{Alvarez2021}, while the data for sticks (PRB2012) have been adapted from Ref.~\onlinecite{Zezelj2012}. The inset shows that, for the fixed value of the number density $n = 2n_c$, the normalized density of the junctions belonging to the backbone of the percolation cluster decreases as the aspect ratio of the discorectangles is increased from 1 to 5, and remains constant within the error bars with further increase. Our results are averaged over 100 independent runs. Since our results obtained using both geometrical and conductive backbones are consistent within the marker size, only one data set is presented in Fig.~\ref{fig:junctions}. All our data collapse to one curve, confirming the proposition of an aspect-ratio-invariant behavior~\cite{Alvarez2021}. For $n/n_c - 1 \gtrapprox 0.2$, and our results for sticks agree the results presented in  Ref.~\onlinecite{Zezelj2012}. Again, a noticeable divergence when $n/n_c - 1 \lessapprox 0.2$ may arise due to both a finite-size effect and a strong dependency on the accuracy of the percolation threshold estimate. However, our results are located significantly above the data for ellipses~\cite{Alvarez2021}. The more natural assumption about differences in methods was not confirmed in our discussion with one of the authors of Ref.~\onlinecite{Alvarez2021}.

\begin{figure}[!htb]
  \centering
  \includegraphics[width=\columnwidth]{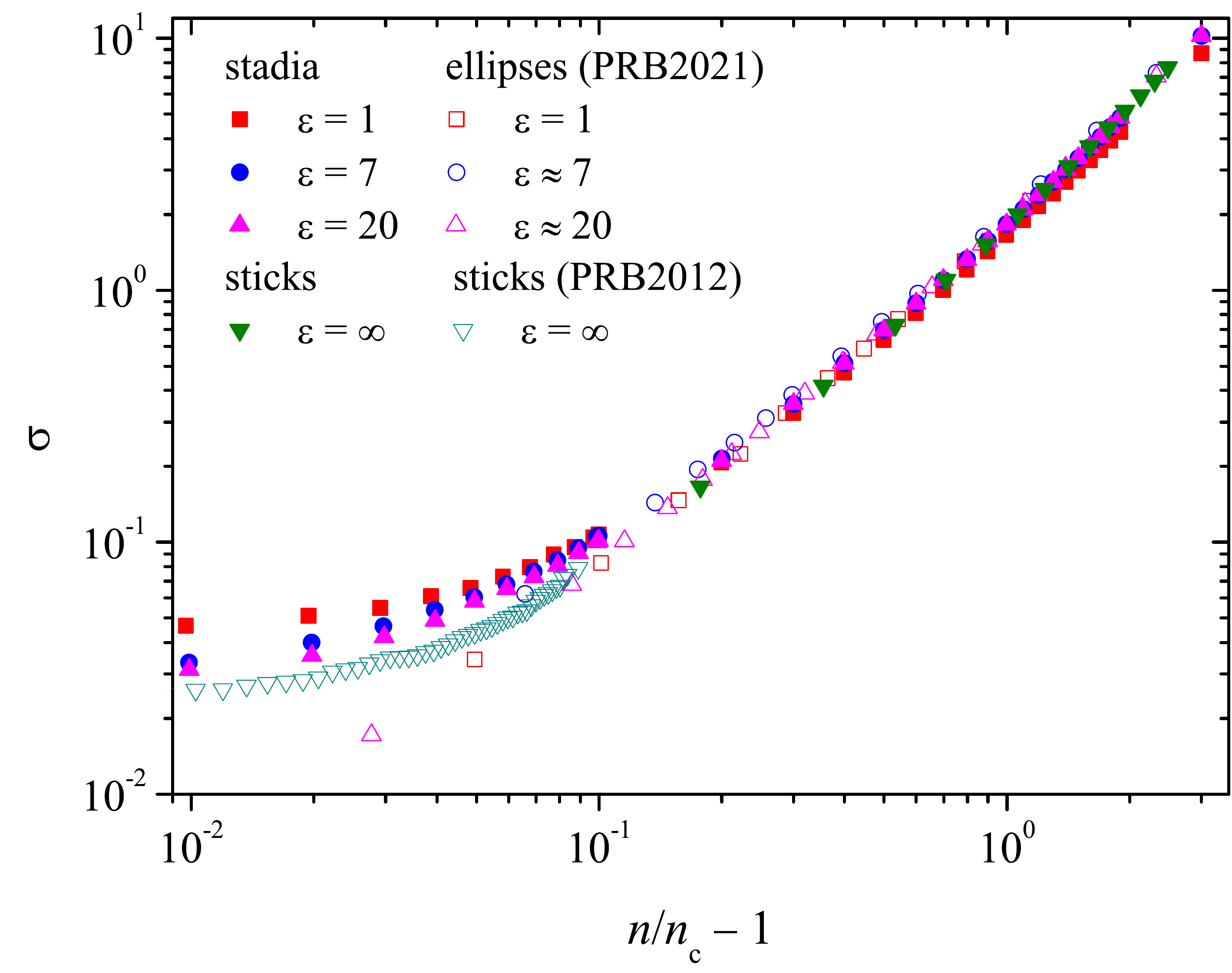}
  \caption{Electrical conductivity against the normalized number density, $n/n_\text{c} - 1$. Filled markers correspond to our results, while open markers correspond to data extracted from the literature. PRB2012 refers to  Ref.~\onlinecite{Zezelj2012}, PRB2021 refers to Ref.~\onlinecite{Alvarez2021}. Our results are averaged over 100 independent runs.}\label{fig:conductivity}
\end{figure}

Figure~\ref{fig:conductivity} compares the behavior of the electrical conductivity for ellipses (filled markers)~\cite{Alvarez2021} and discorectangles (open markers). Again, the dependence of the electrical conductivity on the normalized number density seems to be independent of the aspect ratio. Moreover, the dependencies for both the ellipses~\cite{Alvarez2021} and the discorectangles collapse to single curve except for a region slightly above the percolation threshold.

\begin{figure}[!htb]
  \centering
  \includegraphics[width=\columnwidth]{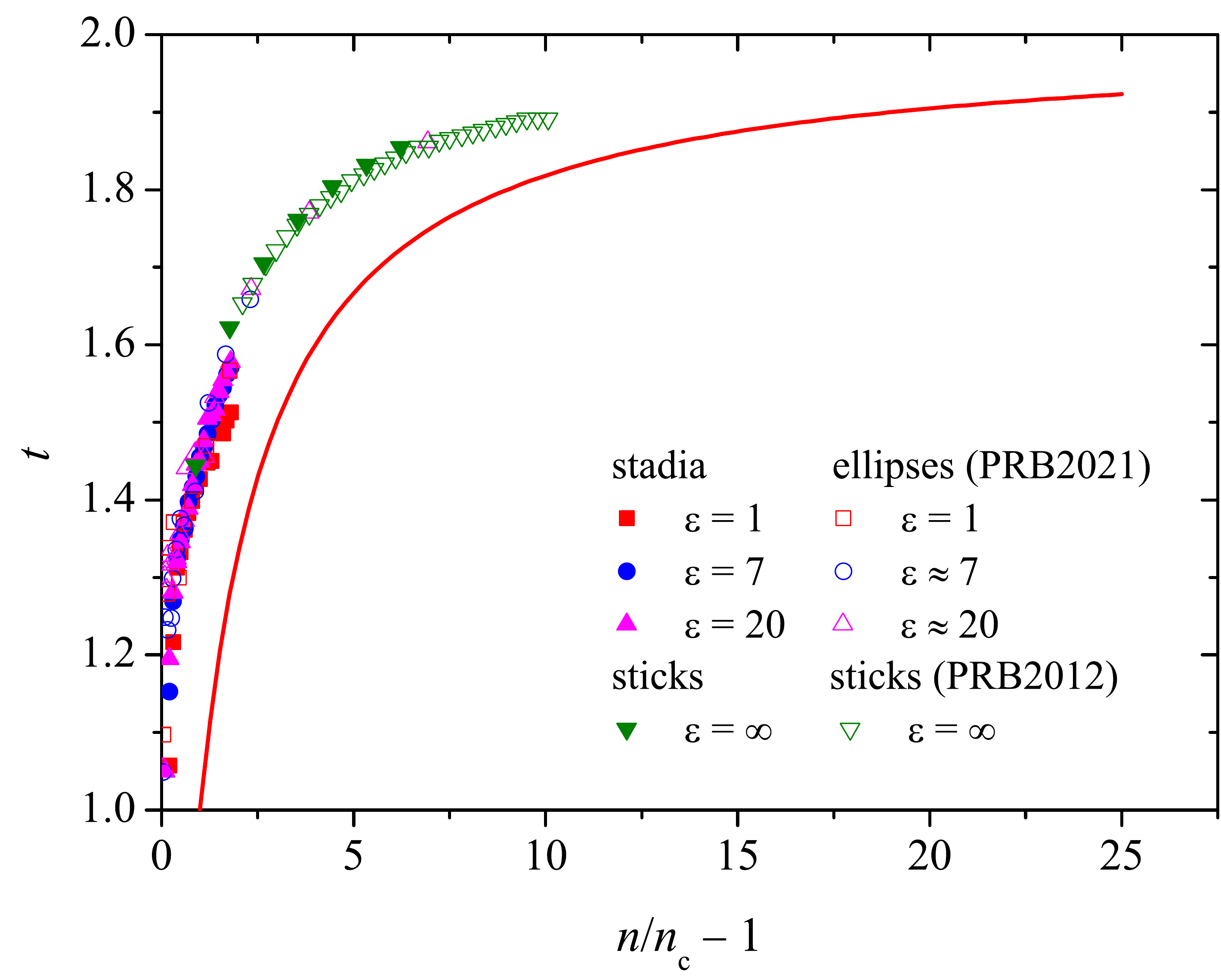}
  \caption{Local transport exponent against the normalized number density, $n/n_\text{c} - 1$. Filled markers correspond to our results, while open markers correspond to data extracted from the literature. PRB2012 refers to  Ref.~\onlinecite{Zezelj2012}, PRB2021 refers to Ref.~\onlinecite{Alvarez2021}. The solid curve corresponds to Eq.~\eqref{eq:tMFAsticks}. Our results are averaged over 100 independent runs.}\label{fig:exponent}
\end{figure}

Figure~\ref{fig:exponent} presents the behavior of the local transport exponent for ellipses (filled markers)~\cite{Alvarez2021} and discorectangles (open markers). Again, the dependence of the electrical conductivity on the normalized number density seems to be independent of the aspect ratio. Moreover, the dependencies for both ellipses~\cite{Alvarez2021} and discorectangles collapse to single curve. The exponent tends to the analytical prediction \eqref{eq:tMFAsticks} when the number density increases.

\section{Conclusion}\label{sec:concl}
Recently, the dynamics of 2D disordered systems of ellipses has been simulated~\cite{Alvarez2021}. Using a particular definition of the normalized proximity to the percolation threshold, the authors found an eccentricity-invariant dynamic behavior. The authors suggested that this invariance might also arise in systems with other particle geometries having zero-width sticks as the limiting case. To check this suggestion, we performed a study using discorectangles including their limiting cases, viz., disks and zero-width sticks. Our computer simulations demonstrate that some properties of random isotropic 2D systems of conductive discorectangles are insensitive to the aspect ratio of the particles. Our study presents some arguments that the suggestion may be correct. At least, the behavior of 2D systems of randomly-placed, conductive, permeable discorectangles is fairly close to that reported for ellipses~\cite{Alvarez2021}. The only exception was the density of the junctions in the backbone. Although our results differ from the results reported in Ref.~\cite{Alvarez2021}, in the limit case of sticks, they do coincide with the results reported in Ref.~\cite{Zezelj2012}. We suggest this deviation is due to differences in the definitions or algorithms. Unfortunately, our conversation with one of the authors of  Ref.~\cite{Alvarez2021} did not elucidate any possible source of this deviation. In order to assist readers, we have presented a detailed description of our own algorithm.

Although, slightly above the percolation threshold, the finite-size effect may presumably be significant, as thw wide range of the number densities, when the invariant behavior can be observed, implies the invariant behavior is insensitive to the domain size. Moreover, we suppose that the reported ``eccentricity-invariant dynamic behavior'' may be observed for a wide range of particles, not only for particles  having zero-width sticks as their limiting case.

\begin{acknowledgments}
The authors acknowledge funding from the Foundation for the Advancement of Theoretical Physics and Mathematics ``BASIS'', grant~20-1-1-8-1.
The authors would also like to thank  A.G.Gorkun for technical assistance, R.K.Akhunzhanov for discussions, and A.~\'Alvarez-\'Alvarez for explanations of some technical details of the study~\cite{Alvarez2021}.
\end{acknowledgments}

\bibliography{DR}

\end{document}